\documentclass[sigconf]{acmart}
\usepackage{multirow}
\usepackage{multirow}
\usepackage{colortbl}
\usepackage{arydshln}
\usepackage[T1]{fontenc}
\usepackage{graphicx}
\usepackage{multirow}
\usepackage{tabularx}
\usepackage{xcolor}
\usepackage{colortbl}
\usepackage{algpseudocode}
\usepackage{algorithm}
\usepackage{bbding}
\usepackage{fontawesome}

\AtBeginDocument{%
  \providecommand\BibTeX{{%
    \normalfont B\kern-0.5em{\scshape i\kern-0.25em b}\kern-0.8em\TeX}}}

\copyrightyear{2024}
\acmYear{2024}
\setcopyright{rightsretained}
\acmConference[SIGIR '24]{Proceedings of the 47th International ACM SIGIR Conference on Research and Development in Information Retrieval}{July 14--18, 2024}{Washington D.C., USA} \acmBooktitle{Proceedings of the 47th International ACM SIGIR Conference on Research and Development in Information Retrieval (SIGIR '24), July 14--18, 2024, Washington D.C., USA} \acmDOI{10.1145/xxxxxx.xxxxxx}
\acmISBN{978-1-4503-XXXX-X/18/06}
\makeatother

\acmConference[SIGIR '24]{}{July 14--18,
  2024}{Washington D.C., USA}

\acmPrice{15.00}
\acmISBN{978-1-4503-XXXX-X/18/06}

\begin{document}

\title{\textit{CFIR:} Fast and Effective 
Long-Text To Image Retrieval for Large Corpora}

\author{Zijun Long} 
\affiliation{
  \institution{University of Glasgow}\streetaddress{}\city{}\country{}}
\email{z.long.2@research.gla.ac.uk}
\author{Xuri Ge$^{\dagger}$}
\affiliation{
\institution{University of Glasgow}\streetaddress{}\city{}\country{}}
\email{x.ge.2@research.gla.ac.uk}
\author{Richard McCreadie}\affiliation{
\institution{University of Glasgow}\streetaddress{}\city{}\country{}}
\email{richard.mccreadie@glasgow.ac.uk}
\author{Joemon M Jose}\affiliation{
\institution{University of Glasgow}\streetaddress{}\city{}\country{}}
\email{joemon.jose@glasgow.ac.uk}
\thanks{$\dagger$ Corresponding author.}

\renewcommand{\shortauthors}{Zijun Long and Xuri Ge, et al.}
\begin{abstract}
Text-to-image retrieval aims to find the relevant images based on a text query, which is important in various use-cases, such as digital libraries, e-commerce, and multimedia databases. Although Multimodal Large Language Models (MLLMs) demonstrate state-of-the-art performance, they exhibit limitations in handling large-scale, diverse, and ambiguous real-world needs of retrieval, due to the computation cost and the injective embeddings they produce. This paper presents a two-stage Coarse-to-Fine Index-shared Retrieval (CFIR) framework, designed for fast and effective large-scale long-text to image retrieval. The first stage, Entity-based Ranking (ER), adapts to long-text query ambiguity by employing a multiple-queries-to-multiple-targets paradigm, facilitating candidate filtering for the next stage. The second stage, Summary-based Re-ranking (SR), refines these rankings using summarized queries. We also propose a specialized Decoupling-BEiT-3 encoder, optimized for handling ambiguous user needs and both stages, which also enhances computational efficiency through vector-based similarity inference. Evaluation on the AToMiC dataset reveals that CFIR surpasses existing MLLMs by up to 11.06\% in Recall@1000, while reducing training and retrieval times by 68.75\% and 99.79\%, respectively. 
We will release our code to facilitate future research at \url{https://github.com/longkukuhi/CFIR}.
\end{abstract}




\begin{teaserfigure}
    \centering
  \includegraphics[width=0.9\textwidth]{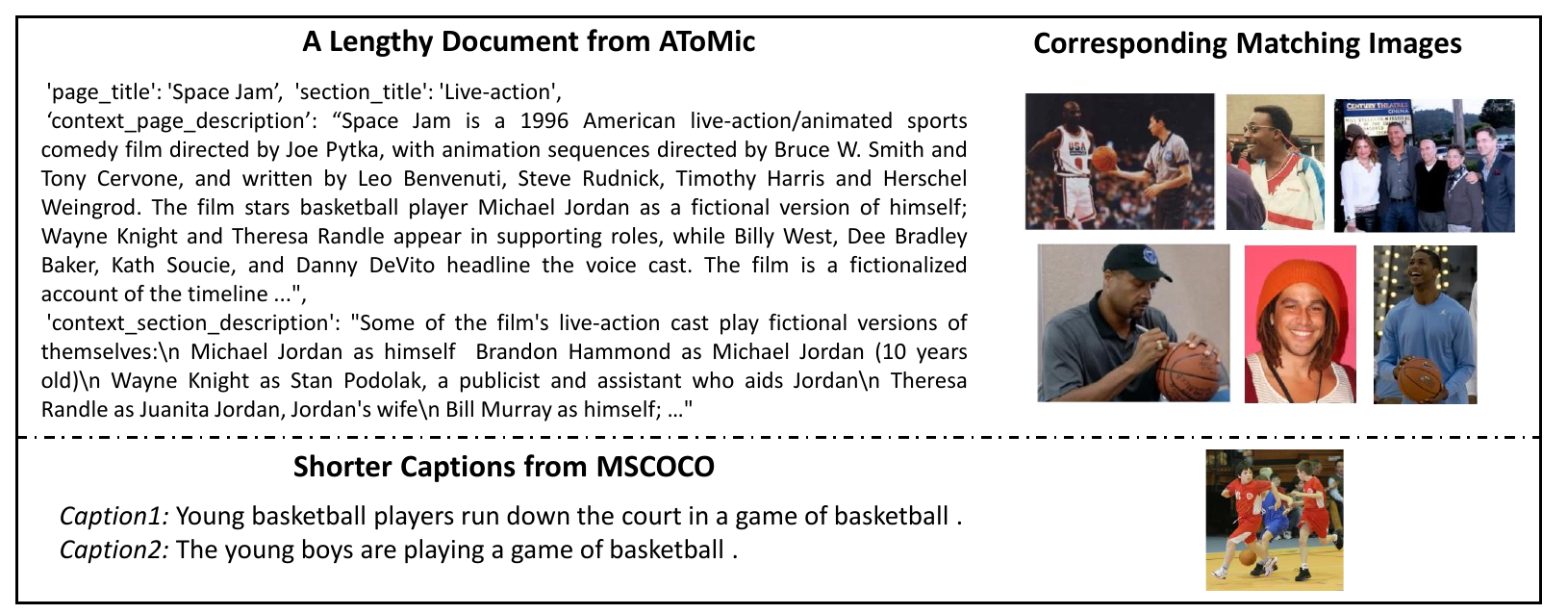}
    \vspace{-1em}
  \caption{Comparison of examples from AToMiC and MSCOCO datasets.}
  \label{fig:motivation1}
\end{teaserfigure}


\maketitle

\begin{figure*}[t]
    \includegraphics[width=0.75\textwidth]{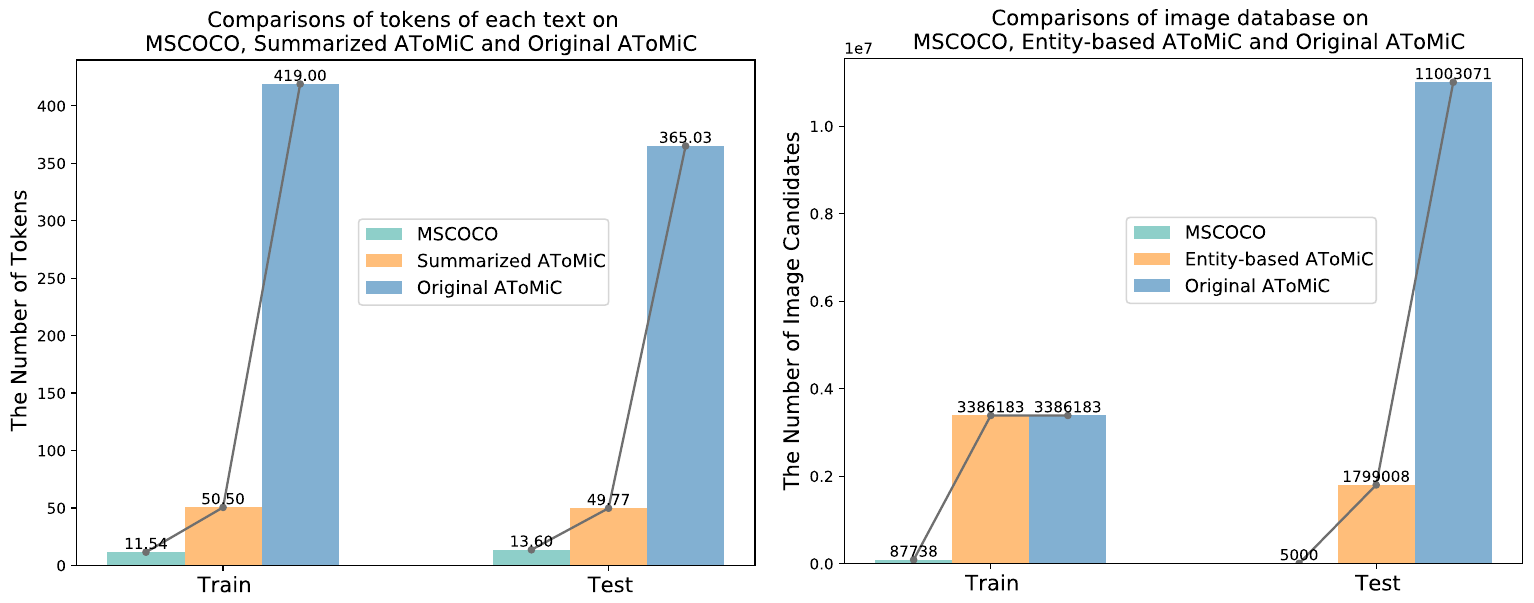}
    \caption{The left is a plot of the average text tokens between MSCOCO short sentences, AToMiC long documents and their summaries and the right is the number of training and testing images of MSCOCO and AToMiC.} 
    \label{fig:motivation2}
    \end{figure*}

\section{Introduction}
Text-to-image retrieval aims to locate relevant images in a database given a text query, which has a wide range of use-cases such as digital libraries \cite{kennedy2005automatic}, e-commerce \cite{wu2021partially}, and multimedia databases \cite{yoshitaka1999survey,DBLP:conf/sigir/HuGCWY23}. Consequently, there is a growing interest in developing effective models for this task. Current state-of-the-art methods predominantly employ Multimodal Large Language Models (MLLMs) such as BEiT-3~\cite{wang2023image} and BLIP~\cite{li2022blip}. These models generate embeddings for both visual and textual inputs, mapping them into a shared space. The mapping function is usually designed to be injective, facilitating a one-to-one correspondence between an instance and its point in the embedding space. Fine-tuning these MLLMs on smaller image-caption datasets such as MSCOCO~\cite{lin2014microsoft} and Flickr30K~\cite{young2014f30k} enables the models to achieve high accuracy in text-to-image retrieval tasks.

However, MLLMs-based methods face limitations particularly in the context of real-world use-cases that involve large-scale, diverse, and ambiguous data such as that illustrated in in Figure \ref{fig:motivation1} and Figure \ref{fig:motivation2}. First, MLLMs-based methods often ignore efficiency concerns. Their model-based similarity inference methods \cite{DBLP:conf/sigir/HongJLWCC21} are computationally demanding, requiring encoding between each query vector and image embedding when ranking. This can result in a computation time of up to 22 hours for a single inference for a large test set \cite{yang2023atomic}, limiting their utility in large-scale retrieval applications despite their high accuracy. Second, real-world use-cases often involve complex queries and images with multiple objects \cite{srinivasan2021wit,yang2023atomic,song2019polysemous}. This contrasts sharply with the comprehensive but short captions found in datasets such as MSCOCO~\cite{lin2014microsoft} and Flickr30K~\cite{young2014f30k}. The nature of this complexity undermines the effectiveness of injective embeddings, which attempt to map diverse meanings/senses to a single point in shared space, which could be an inaccurate weighted geometric mean of all the desirable points \cite{song2019polysemous}. This is particularly problematic in long-text query to image retrieval tasks, where accumulated ambiguities significantly hinder the performance of Multimodal Large Language Models (MLLMs) \cite{yang2023atomic}.  Third, injective embeddings struggle with partial text-to-image associations \cite{song2019polysemous}. In a long query, only a subset of sentences may relate to specific regions or aspects of an image, while the rest discuss unrelated subjects. Additionally, a single sentence may describe just a particular region of an image rather than its entirety.

To address these challenges, this paper presents a novel two-stage Coarse-to-Fine Index-shared Retrieval (CFIR) framework, jointly optimizing effectiveness and efficiency. The first stage is entity-based Ranking (ER) and the ER result is used to construct a shared entity-based image candidates index. ER is designed to be computationally cheap, using pre-computed image embeddings from a cache. By replacing the entire document with a representation comprising its entities as the query, we transform the retrieval task from one query to one target, to multiple queries to multiple targets, accommodating the ambiguity inherent in long documents and images. This transformation makes ER well-suited for use-cases demanding relevance but not exact matching, such as multimedia content creation~\cite{deldjoo2020recommender,RN155}, where a diverse array of images is beneficial for illustrative purposes. Furthermore, ER can be used to filter out the majority of irrelevant candidates prior to the re-ranking stage, thereby reducing the overhead from the more powerful encoder used in the re-ranking stage. The second stage is Summary-based Re-ranking (SR). By summarizing long documents as queries and using entity-based image candidates from the pre-computed shared index, SR further mitigates ambiguity, making the framework robust against partial text-to-image associations and reducing encoding time. The main contributions of this work are as follows:

\begin{enumerate}
\item We introduce the two-stage Coarse-to-Fine Index-shared Retrieval (CFIR) framework to tackle the effectiveness and efficiency challenges of state-of-the-art MLLMs based approaches in the more challenging real-world scenario, with Entity-based and Summary-based Ranking stages. 
\item We introduce a novel Decoupling-BEiT-3 encoder optimized for both ER and SR stages. This encoder employs a decoupled encoding design for vector-based distance computation, enhancing both training and retrieval efficiency.  
\item CFIR is evaluated on the AToMiC dataset, showing an 11.06\% improvement in Recall@1000 and reducing computational times by 68.75\% and 99.79\% in training and retrieval, respectively.
\end{enumerate}

\section{Related Work}
\label{sec:task}

\subsection{Small-scale and Caption-based Text-to-Image Retrieval}
A majority of existing Text-to-Image retrieval methods \cite{huang2018learning,li2019visual,ge2021structured,IMRAM,ge2023cross,Gradual,ge20243shnet,long2024multiway} concentrate on small-scale and caption-based benchmarks, such as MSCOCO \cite{lin2014microsoft} and Flickr30K \cite{young2014f30k}. They often excel by employing intra-modal and inter-modal attention mechanisms to align entity semantics across modalities. Specifically, they try to ensure that the meaning or representation of specific entities is consistent when interpreted through different modalities. Methods such as \cite{feng2014cross,wang2018joint,shao2019two,wang2019matching,qu2023learnable} adopt a two-stage retrieval strategy to further refine the feedback results based on one-stage ranking to obtain more accurate retrieval. For instance, MTFN \cite{wang2019matching} introduced a generic text-to-image re-ranking scheme for refinement during the inference process without requiring additional training procedures. Moreover, JGCAR \cite{wang2018joint} and LeaPRR \cite{qu2023learnable} proposed modeling the higher-order neighbor relationship-aware attentions for text-image retrieval in a learnable two-stage re-ranking paradigm. However, most of these methods designed a relatively complex first-stage multi-modal interaction model to establish a precise candidate set for the subsequent stage re-ranking process. These methods are computationally intensive and do not scale well to large datasets with long textual queries and diverse topics of images. Recent advancements in Multimodal Large Language Models (MLLMs) \cite{radford2021learning,cherti2023reproducible,li2022blip,singh2022flava,wang2023image,DBLP:conf/sigir/TianWXCSS22,DBLP:conf/sigir/ZhaoGH0YL23,DBLP:conf/sigir/LinJSLSN23,long2023robollm} signify a paradigm shift in the field. While these models offer robust performance in Text-to-Image retrieval, their application to large-scale, long-text queries for image retrieval presents challenges in both efficiency and effectiveness due to the computational cost of MLLMs and issues with injective embeddings \cite{song2019polysemous,yang2023atomic}. In response, this paper proposes a novel two-stage coarse-to-fine index-shared retrieval framework tailored to address these challenges.

\vspace{-1mm}
\subsection{Large-scale Text-to-Image Retrieval datasets}
\label{sec:atomic}
AToMiC \cite{yang2023atomic} is a recently released dataset for large-scale long-text to image retrieval, introduced by TREC \footnote{https://trec.nist.gov}. AToMiC is built upon the WIT dataset \cite{yang2023atomic}. AToMiC distinguishes itself by focusing on section-level image-text associations for multimedia content creation, emphasizing the use of English Wikipedia sections without images for a more realistic text-image context. Unlike WIT, which is employed for broader tasks like image-caption matching and generation, AToMiC is tailored for ad hoc retrieval tasks, reusing WIT's images and metadata but providing image pixel values in a standardized format. Therefore, AToMiC is the only and the best option to evaluate the performance of models in the context of large-scale long-text to image retrieval. evaluate long In this paper, we target the large-scale long-text to image retrieval (LLIR) task in AToMiC. This task is designed to retrieve images from large image collections based on a long-text query for scenarios such as article writing. It encompasses more than 21 million images and textual documents, offering two distinct evaluation settings: a base setting and a large setting.

Both settings utilize a training set comprising 4,401,903 query-document relevance assessments (qrels), a validation set with 17,801 qrels, and a test set containing 9,873 qrels. In the base setting, each image is accompanied by at least one corresponding long document, and retrieval candidates are limited to the labelled images that are as least relevant to one document. In contrast, the large setting extends the candidate pool by incorporating an additional 7,608,283 images, offering a more challenging retrieval context.

\begin{figure*} 
\includegraphics[width=1\textwidth]{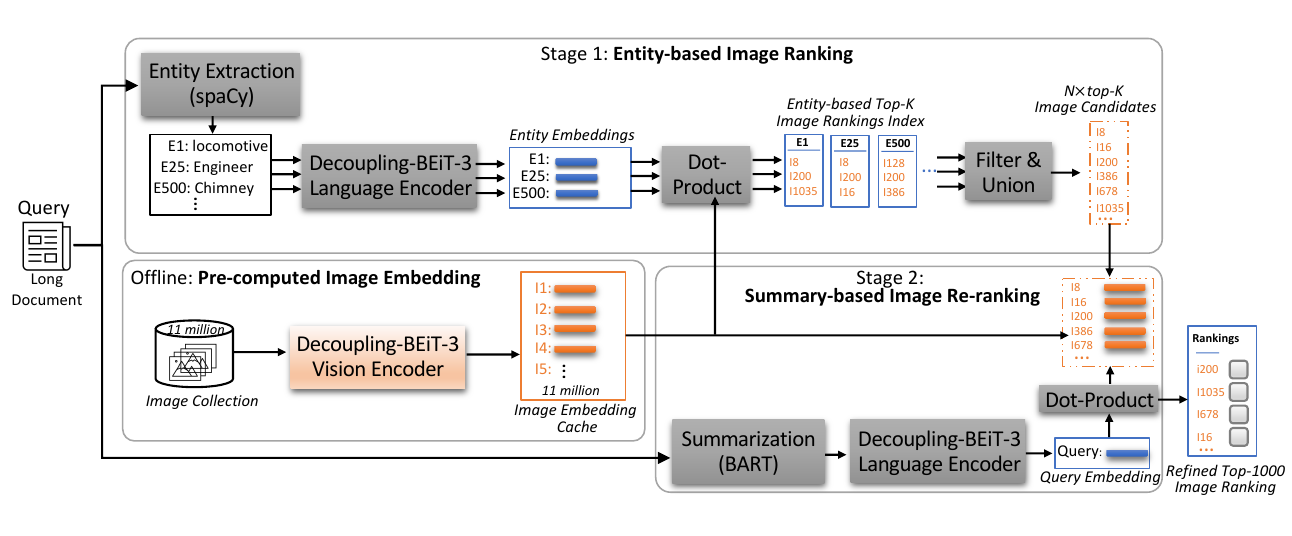}
\vspace{-2em}
\caption{The overall architecture of the proposed CFIR for large-scale document-to-image retrieval. } \label{Framework}
\end{figure*}
\subsection{Challenges in MLLM-based approaches}
\label{sec:LLIRchallenges}
In contrast to small-scale image-caption datasets such as MS COCO \cite{lin2014microsoft}, as shown in Figure \ref{fig:motivation2}, which comprises 165,000 images with text descriptions averaging 11.53 tokens and corresponding to a single ground-truth image, AToMiC presents a more realistic simulation of LLIR applications. It uses longer, multi-faceted (ambiguous) real-world documents, averaging 415 text tokens, and maps them to multiple ground-truth images. The characteristics of LLIR present challenges for state-of-the-art MLLM-based approaches, primarily concerning retrieval effectiveness and computational efficiency, due to their reliance on a robust visual encoder \cite{long157,long2024crisisvit}.

In regard to retrieval effectiveness, the issues manifest in two ways. First, the complexity and multi-faceted nature of long documents introduce semantic ambiguities, making it more difficult for injective models to accurately discern text-to-image similarities. Second, this challenge is further exacerbated by the expanded pool of candidate images, complicating the task of identifying the most relevant matches.

On the computational front, inefficiencies are also divided into two parts. Firstly, the inference stage in current MLLM-based methods demands an exhaustive pairing of each query with database items, which are then processed by the MLLM to predict matching scores~\cite{DBLP:journals/corr/abs-1908-06066,DBLP:journals/corr/abs-1908-03557,DBLP:conf/eccv/Li0LZHZWH0WCG20,DBLP:conf/nips/LuBPL19,DBLP:conf/emnlp/TanB19,RN150}. This model-based similarity inference is both computationally intensive and time-consuming, particularly when compared to vector-based distance computations. Moreover, contrastive learning techniques \cite{RN152,RN156} are often employed to enhance intra-modal alignment, thereby improving effectiveness. However, this approach increases the computational burden. Secondly, the act of encoding long documents for semantic matching is itself a time-consuming process. These inefficiencies limit the practical applicability of MLLMs to large-scale retrieval tasks, despite their promising accuracy.

\section{Methodology}

To systematically address the challenges inherent in Multimodal Large Language Models (MLLMs) for Large-Scale Long-Text to Image Retrieval (LLIR), we propose a two-stage coarse-to-fine index-shared retrieval (CFIR) framework, as shown in Figure \ref{Framework}. The pseudocode for the corresponding training algorithm is shown in Figure \ref{algorithm:training}. Moreover, the pseudocode for the corresponding retrieval algorithm (during testing) is shown in Figure \ref{algorithm:testing}. CFIR is subdivided into two core stages:  Entity-based Ranking (ER) and  Summary-based Re-ranking (SR). We also introduce a novel Decoupling-BEiT-3 encoder optimized for both ER and SR stages.

\begin{figure}
    \begin{algorithmic}[1] 
\Require 
\ Long-text set $\mathcal{D}$,  Image set $\mathcal{I}$, 
\ A pre-trained version of our proposed decoupling-BEiT-3 model,
\ Text-entity extractor spaCy,
\ Text-summary generator BART large model \cite{lewis2020bart}.
\Ensure 
\ Entity-based image ranking index $\mathcal{E} _{index}$, 
\ Image embedding index $\mathcal{V} _{index}$.

\noindent \textbf{// Stage 0: Pre-computing image embedding index.}

\For{for each image in $\mathcal{I}$ do}

\State
Encode the image using our proposed D-BEiT-3 model with image expert to generate embedding $v_{i}$; 
\State
$\mathcal{V} _{index}$.append($v_{i\rightarrow index}$);
\EndFor

\noindent \textbf{// Stage 1: Entity-based Ranking (ER).}
\For{each long-text query in  $\mathcal{D}$}
    \State
    Extract entities $\{e_1, \ldots, e_N\}$ of $i$-index long-text query by spaCy; 
        \State 
        Encode $e_j$ using our proposed D-BEiT-3 model with text encoder to generate embedding $t_{j}$; 
        \State
        Build a similarity score list $S$;
        
    \For{for each image embedding in $\mathcal{V}$ do}
        \State
        Compute the similarity (dot-product) between $t_{j}$ and the selected image embedding;
        $\mathcal{S} _{index}$.append($t_{j\rightarrow index}$);
    \EndFor
    \State
    Choose Top-K from $S$ to build $\mathcal{E}_{i\rightarrow index}$ = list($[I_1, \ldots,I_K]$);

\EndFor 

\noindent\textbf{// Stage 2: Summary-Based Re-ranking (SR).}
\For{each-index long-text query in  $\mathcal{D}$}
    \State 
    Extract the entities from $i$-index long-text query by spaCy;
    \State
    Obtain the corresponding pre-stored Top-K image ranking index from $\mathcal{E} _{index}$ based on the extracted entities;
    \State
     Filter \& Union repeated candidates ranking index form a candidate set;
     \State
    Obtain the corresponding  image embedding set $\mathcal{V}_{candidates}$ from $\mathcal{V} _{index}$;
    
    \State  
    Summary the $i$-index long-text query by the BART large model;
    \State 
    Encode the summary by using our proposed D-BEiT-3 with text expert as $\mathbf{q}_{index}$;
    
    \State
    Compute the similarities (dot-product) between the query embedding $\mathbf{q}_{index}$ and coarse-grained image embedding set $\mathcal{V}_{candidates}$;
    
    \State 
    \Return image ranking.
\EndFor 
\end{algorithmic} 
\caption{CFIR Training Procedure}
\label{algorithm:training}
\vspace{-5mm}
\end{figure}

\begin{figure}
    \begin{algorithmic}[1] 
\Require 
\ long-text query $q$, 
\ A pre-trained version of our proposed decoupling-BEiT-3 model,
\ Text-entity extractor spaCy,
\ Text-summary generator BART large model \cite{lewis2020bart}.
\ Entity-based image ranking index $\mathcal{E} _{index}$, 
\ Image embedding index $\mathcal{V} _{index}$.

\noindent\textbf{// Summary-Based Re-ranking (SR).}

    \State 
    Extract the entities from the long-text query $q$ by spaCy;
    \State
    Obtain the corresponding pre-stored Top-K image ranking index from $\mathcal{E} _{index}$ based on the extracted entities;
    \State
     Filter \& Union repeated candidates ranking index form a candidate set;
     \State
    Obtain the corresponding  image embedding set ${q}_{candidates}$ from $\mathcal{V} _{index}$;
    
    \State  
    Summary the long-text query $q$ by the BART large model;
    \State 
    Encode the summary by using our proposed D-BEiT-3 with text expert as $\mathbf{q}_{encoded}$;
    
    \State
    Compute the similarities (dot-product) between the query embedding $\mathbf{q}_{encoded}$ and  image embedding set ${q}_{candidates}$;
    
    \State 
    \Return image ranking.
\end{algorithmic} 
\caption{CFIR Retrieval (testing) Procedure}
\label{algorithm:testing}
\end{figure}

\subsection{The Proposed Decoupling-BEiT-3}
\label{sec:streamlinedbeit3}

\begin{figure*}
\centering
\includegraphics[width=0.7\textwidth]{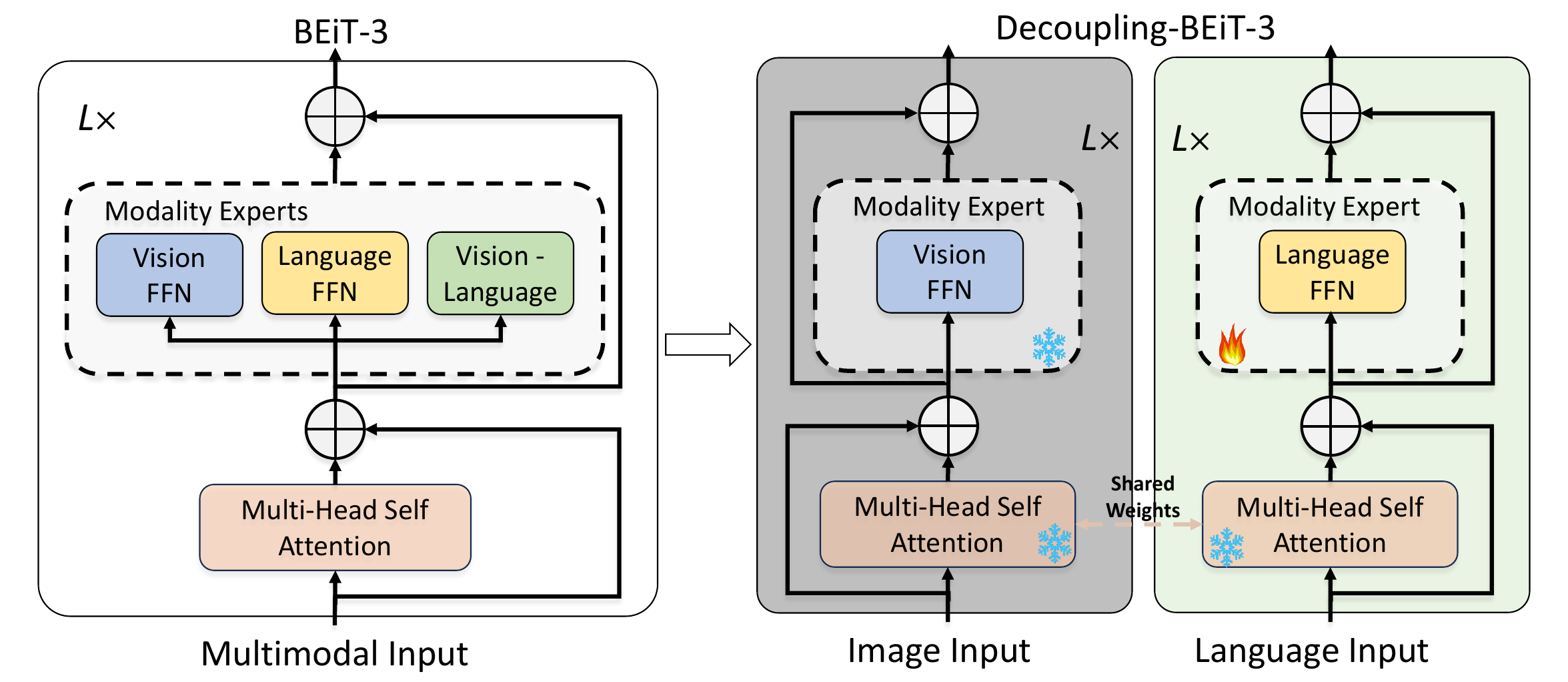}
\caption{The demonstration of differences between the original architecture of BEiT-3 model and our Decoupling-BEiT-3. } \label{fig:Beit3}
\end{figure*}

The BEiT-3 model is originally constructed as a MultiWay Transformer design~\cite{wang2023image}. As depicted on the left side of Figure~\ref{fig:Beit3}, the MultiWay Transformer block in BEiT-3 features shared self-attention modules and a pool of feed-forward networks (i.e., modality experts) tailored for different modalities. 

To better fit the LLIR task, we propose a decoupling-BEiT-3 (D-BEiT-3) as the MLLM encoder in our CFIR. Our D-BEiT-3 architecture removes the Vision-Language (VL) expert, as shown on the right side of Figure~\ref{fig:Beit3}. This design is motivated by three primary considerations. First and most importantly, without the VL expert, we decouple the encoding of visual and text input and transition from model-based similarity inference to vector-based distance computation, which is significantly faster. We also index the image vector to further reduce computational cost during both training and testing. Specifically, if we were to use the VL expert in the original BEiT-3 model, it would be necessary in the inference stage to exhaustively pair the query with each database item and then feed these pairs into the BEiT-3 model to predict matching scores. Second, although BEiT-3's design is effective for accurate instance-level alignment between text and images, it is optimized for image descriptions (captions) that are both precise and comprehensive. This specialization is at odds with the multi-faceted (ambiguous), long documents found in the AToMiC dataset, which often describe multiple images and objects. This inherent ambiguity causes the model to underperform, as identified in \cite{yang2023atomic}. By eliminating the vision-language expert, our architecture better suits the less stringent semantic alignment requirements of the LLIR task. Third, our streamlined architecture results in a 30.4\% reduction in model parameters compared to the original BEiT-3 model, significantly enhancing both training and inference efficiency.

\vspace{-3mm}
\subsection{Entity-based Ranking (ER)}

The Entity-based Ranking (ER) stage serves two primary functions. First, it generates a Top-K ranking of images for each unique named entity extracted from long-text queries, thereby mitigating ambiguity and partial associations. This is achieved through a shift from a one-to-one to a multiple-queries-to-multiple-targets retrieval paradigm. Second, ER effectively prunes irrelevant image candidates, paving the way for the subsequent  re-ranking. To facilitate this and improve efficiency, we construct an entity-based candidate index that maps each entity to its likely corresponding images, based on the Top-K ranking obtained from the ER stage.  Repeated entities across different query documents can be swiftly retrieved from the index, reducing computational costs. Consequently, in the training phase, we construct an entity-based candidate index encompassing all ranking results for entities present in the training samples. If an unknown entity (not included in the training samples) appears during retrieval, it is disregarded. This approach has a negligible impact on performance, given the extensive dataset of over 11 million training samples. Additionally, it significantly boosts efficiency by obviating the necessity to recompute any Entity Retrieval (ER) stage for new entities during test retrieval. This approach allows for the computation of the entity-based ranking to be performed only once, during the training phase. Furthermore, if there is a need to augment system performance by incorporating additional unknown entities into the entity-based candidate index, this can be efficiently achieved offline. This entails appending the ranking results to the index at a later time, rather than conducting this process online during retrieval, which has no impact on the retrieval efficiency.

To accomplish this, as shown in Figure \ref{Framework}, we first employ the advanced Natural Language Processing (NLP) library, spaCy, to extract name entities from each long document. SpaCy's robust entity extraction capabilities serve as an effective mechanism for generating entity queries. Subsequently, we use a pre-trained and frozen D-BEiT-3 to encode these entities, eliminating the need for additional training and thereby enhancing computational efficiency. We then retrieve the Top-\(K\) candidate images based on their similarity scores with each entity. These similarity scores are calculated as the dot-product between each entity's embedding and the embeddings of the image candidates, which are retrieved from a pre-computed shared image embedding cache.

\vspace{-2mm}
\subsection{Summary-Based Re-ranking (SR)}

To achieve precise image matching, we introduce the Summary-based Re-ranking (SR) stage, tailored for LLIR. Contrary to the ER stage, SR focuses exclusively on precise image matching of a document's key information - document summary - to refine the ranking of previously identified entity-based image candidates. Text summaries are generated using BART \cite{lewis2020bart}, chosen for its ability to produce concise summaries that capture the document's core semantics. Notably, BART \cite{lewis2020bart} supports a larger maximum input token count of 1024, accommodating the average token number of 419 in AToMiC queries, compared to BERT's \cite{devlin2019bert} 512-token limit. The summary effectively mitigates the semantic ambiguity and partial association problems in long document query, thus improving the retrieval effect. 

During training, only the language expert component of our D-BEiT-3 is fine-tuned, leaving the remaining modules frozen. This approach facilitates the construction of a pre-computed shared image embedding cache, striking an optimized balance between training efficiency and retrieval efficacy.  The entire training process is optimized by a symmetric cross-entropy loss over the similarity scores between the text representation and image representations.

In inference, we utilize pre-computed image embeddings from a shared cache, negating the need for recalculations in each training epoch. Candidate selection bypasses full-database retrieval, opting for a union subset comprising the top-\(K\) entity-queried candidates for each entity in the query. The theoretical candidate set size should be $N$$\times$Top-$K$, where $N$ is the number of entities in a query, a count significantly lower than that of the comprehensive image database. For instance, with 10 entities and selecting the Top-10,000 results for each, we have an image candidate pool of 100,000. This size is substantially smaller compared to the 4 million images in the AToMiC base setting or the 11 million images in the larger setting. This strategy reduces computational load and is made possible by the shared entity-based candidates index. Moreover, empirical observations indicate a substantial overlap—approximately 45.6\%—between candidate sets for distinct entities, resulting in an actual filtered set size substantially smaller than $N$$\times$Top-$K$. Consequently, it is only necessary to calculate the dot-product between the embeddings of filtered image candidates and the summary query embeddings to establish the final ranking. This approach eliminates the need for computations involving all images in the database.

\section{Experimental Setup}
We utilize the Base (H = 768) and Large model (H = 1024) of our streamlined version of BEiT-3 as our encoder in CFIR, denoted as CFIR-B and CFIR-L respectively, where H is the hidden size. Additionally, we include comparisons with two state-of-the-art Multimodal Large Language models: our proposed D-BEiT-3 and OpenCLIP \cite{cherti2023reproducible}. OpenCLIP is an open-source variant of OpenAI's CLIP, specialized for multi-modal learning with text and images. It enables zero-shot classification and cross-modal retrieval in a shared embedding space without requiring task-specific fine-tuning.

We adhere to the experimental setup for BEiT-3 \footnote{https://github.com/microsoft/unilm/tree/master/beit3}, and fine-tune for the AToMiC dataset. For OpenCLIP we include the results reported in \cite{yang2023atomic}. We conduct training over 30 epochs. For image data augmentation in training, we employ AutoAugment~\cite{DBLP:conf/cvpr/CubukZMVL19}. Throughout all fine-tuning experiments, we choose the Adam optimizer with a learning rate set at \(1\times10^{-4}\), a weight decay of \(0.05\), and a batch size of \(512\). We also integrate a dropout rate of 0.1.  

Adhering to creator of AToMiC dataset \cite{yang2023atomic} and to ensure a fair comparison, we assess the performance of all methods using established metrics: recall at 1000 (R@1000) and mean reciprocal rank at 10 (MRR@10). Additionally, we report both training and retrieval times to evaluate model efficiency. Our complete code is available at \url{https://anonymous.4open.science/r/CFIR-B7EE/}, which will be made publicly available upon acceptance.

\section{Experiments}

This section aims to evaluate the efficacy and efficiency of our proposed CFIR framework in addressing the challenges outlined in Section \ref{sec:LLIRchallenges}. Our evaluation encompasses both the base and large settings of the AToMiC dataset, as discussed in Section \ref{sec:atomic}. Specifically, we answer the following research questions.
\begin{itemize}
    \item RQ1: What is the Benefit and Cost of Freezing the Image Encoder?
    \item RQ2: How does CFIR perform compared to state-of-the-art approaches?
    \item RQ3: How does the CFIR model demonstrate scalability in the context of the larger and more challenging AToMiC Large Setting?
\end{itemize}

\subsection{What is the Benefit and Cost of Freezing the Image Encoder? (RQ1)}

We evaluate the computational and performance trade-offs of freezing the image encoder by comparing its performance with that of whole-model fine-tuning. The comparison focuses on two models: OpenCLIP and D-BEiT-3, as detailed in upper block of Table \ref{tab1} for AToMiC base setting and Table \ref{tab2} for AToMiC large setting. The primary motivation for freezing the image encoder lies in the creation of an image embedding cache, which substantially mitigates computational overhead on encoding images during both training and retrieval phases.

In the AToMiC base setting, when compared to the fully fine-tuned OpenCLIP large model (OpenCLIP-L-Full), D-BEiT-3 large model with frozen image encoder (D-BEiT-3-L-Frozen) demonstrates a modest drop of 3.48\% in Recall@1000, making the performance loss acceptable given that OpenCLIP-L was the previous state-of-the-art model. Moreover, D-BEiT-3-L-Frozen experiences a decrease in performance, with a 6.03\% drop in Recall@1000 and a 0.02 reduction in MRR@10, compared to its whole-model fine-tuned version (D-BEiT-3-L-Full). 
However, D-BEiT-3-L-Frozen is fine-tuned with only 56\% of the parameters and achieves a 30\% reduction in training time compared to the OpenCLIP-L-Full. Furthermore, the implementation of an image embedding cache leads to a substantial reduction in retrieval time.  The query time decreases by 81\% for D-BEiT-3-L (from 2257.5 ms to 425.1 ms) and by 77.8\% for D-BEiT-3 base model (D-BEiT-3-B) (from 1640.7 ms to 363.6 ms) when transitioning from fully D-BEiT-3-L/B-Full to D-BEiT-3-L/B-Frozen. This leads to a considerable reduction in training time. For instance, D-BEiT-3-L-Frozen shows a decrease of 23 hours per epoch compared to D-BEiT-3-L-Full, while D-BEiT-3-B-Frozen exhibits an 8-hour reduction per epoch relative to D-BEiT-3-B-Full. This is particularly significant as model testing forms an integral part of the training process. Employing a whole-model fine-tuning approach would necessitate approximately 2280 hours (around 95 days) to train the model over 30 epochs on a single GPU. Such a duration is impractical in typical academic experimental environments.

As for the more challenging scenario, AToMiC Large setting which has 11 millions long-text and 11 millions images, D-BEiT-3-L-Frozen demonstrates remarkable efficiency. As for another important aspect in evaluating the efficiency, the retrieval latency, D-BEiT-3-L-Frozen and D-BEiT-3-B-Frozen needs just 17.9\% and 22.4\% of the retrieval time required by D-BEiT-3-L-Full and D-BEiT-3-B-Full. This significant reduction in retrieval time underscores the enhanced scalability and aptness of these models for large-scale applications.

In summary, we construct an shared image embedding cache and a shared entity-based image ranking index to markedly enhance training and retrieval efficiency, predicated on the freezing of the image encoder. While there is a performance trade-off, the gains in efficiency render the model highly deployable and make it possible for large-scale text-to-image tasks within an academic budget. This result answers our proposed research question 1 that what is the benefit and cost of freezing the image encoder. Consequently, we choose to build our CFIR framework on this premise.
\vspace{-1.5mm}
\begin{table*}
    \centering
    \caption{Comparisons of experimental results on AToMiC base setting for large-scale document-to-image retrieval. VE and LE indicate the vision encoder and language encoder with (\faFire) or without (\Snowflake) fine-tuning. $\#$ P (M) indicates trainable parameters of the multi-modal encoders. T-t (Hour/epoch) means the training time and R-t (millisecond/query) means the retrieval time for each query. For OpenCLIP runs, - indicates the metric  was not reported in \cite{yang2023atomic}.  }
    \label{tab1}
    \small
    \renewcommand\tabcolsep{5.0pt}
    \begin{tabular}{ccc|c|c|ccc}
    \hline

    \hline
    \multicolumn{1}{c}{\multirow{2}{*}{Method}} & \multicolumn{1}{c}{\multirow{2}{*}{VE}} & \multicolumn{1}{c|}{\multirow{2}{*}{LE}} & \multicolumn{1}{c|}{\multirow{2}{*}{\# P}} & \multicolumn{1}{c|}{\multirow{2}{*}{T-t}} & \multicolumn{3}{c}{AToMiC Base Setting } \\
     &  &  &  &  & R-t &  MRR@10  & R@1000  \\
    \hline
     OpenCLIP-B&  \faFire & \faFire     & 197    &  -  & -   &  0.043    	 &  44.68 \\ 
     D-BEiT-3-B (proposed model)  &   \faFire &  \faFire   & 155    &  16  & 1640.7   &  0.048 	 &  50.65 \\ \hdashline
     OpenCLIP-L & \faFire & \faFire     & 645    & -   & -   &  0.065   & 54.84  \\ 
    D-BEiT-3-L (proposed model) & \faFire & \faFire        & 490    & 76   & 2257.5    &  0.085   & 57.39  \\ 
    \hline
    
    \hline
     OpenCLIP-B   &\Snowflake &\faFire &  110   & -    & -  &  0.037   &  39.66   \\
      D-BEiT-3-B (proposed model) &\Snowflake & \faFire  &   87   &  8   & 363.6  &  0.042    	 &  43.28   \\
    \cellcolor{gray!10}\textbf{CFIR-B (proposed framework)}  & \cellcolor{gray!10}\Snowflake & \cellcolor{gray!10}\faFire  & \cellcolor{gray!10}87  & \cellcolor{gray!10} \textbf{5}  & \cellcolor{gray!10}\textbf{4.2}   &  \cellcolor{gray!10}\textbf{0.052}   & \cellcolor{gray!10}\textbf{50.72} \\
    \hdashline
     OpenCLIP-L & \Snowflake & \faFire   & 340   & - & -   &  0.063    & 50.13   \\
     D-BEiT-3-L (proposed model) & \Snowflake & \faFire    & 305     &  53 & 425.1   &  0.065  & 51.36  \\
    \cellcolor{gray!10}\textbf{CFIR-L (proposed framework)}  & \cellcolor{gray!10}\Snowflake & \cellcolor{gray!10}\faFire  & \cellcolor{gray!10}305  & \cellcolor{gray!10} \textbf{45}   & \cellcolor{gray!10} \textbf{4.7} &  \cellcolor{gray!10}\textbf{0.081}   & \cellcolor{gray!10}\textbf{55.68} \\
    \hline
    
    \hline
    \end{tabular}
    \end{table*}

\begin{table*}
    \centering
    \caption{Comparisons of experimental results on large setting for large-scale document-to-image retrieval.}
    \label{tab2}
    \small
    \renewcommand\tabcolsep{5.0pt}
    \begin{tabular}{ccc|c|c|ccc}
    \hline

    \hline
    \multicolumn{1}{c}{\multirow{2}{*}{Method}} & \multicolumn{1}{c}{\multirow{2}{*}{VE}} & \multicolumn{1}{c|}{\multirow{2}{*}{LE}} & \multicolumn{1}{c|}{\multirow{2}{*}{\# P}} & \multicolumn{1}{c|}{\multirow{2}{*}{T-t}} & \multicolumn{3}{c}{AToMiC Large Setting }     \\
     &  &  &  &  & R-t &  MRR@10  & R@1000  \\
    \hline
     D-BEiT-3-B (proposed model)  &   \faFire &  \faFire   & 155    &  16   & 6016.3 &  0.019 &  37.11\\ \hdashline
    D-BEiT-3-L (proposed model) & \faFire & \faFire        & 490    & 76     & 8140.2   & 0.038 & 43.37\\ 
    \hline
    
    \hline
      D-BEiT-3-B (proposed model) &\Snowflake & \faFire  &   87   &  8     & 1349.1  & 0.015 & 36.16\\
    \cellcolor{gray!10}\textbf{CFIR-B (proposed framework)}  & \cellcolor{gray!10}\Snowflake & \cellcolor{gray!10}\faFire  & \cellcolor{gray!10}87  & \cellcolor{gray!10} \textbf{5}  & \cellcolor{gray!10} \textbf{364.1} & \cellcolor{gray!10}\textbf{0.021}		& \cellcolor{gray!10}\textbf{38.07}\\
    \hdashline

     D-BEiT-3-L (proposed model) & \Snowflake & \faFire    & 305     &  53  & 1458.5 &0.026	 &39.36\\
    \cellcolor{gray!10}\textbf{CFIR-L (proposed framework)}  & \cellcolor{gray!10}\Snowflake & \cellcolor{gray!10}\faFire  & \cellcolor{gray!10}305  & \cellcolor{gray!10} \textbf{45}   & \cellcolor{gray!10} \textbf{364.6} & \cellcolor{gray!10}\textbf{0.030}	 & \cellcolor{gray!10}\textbf{42.51}\\
    \hline
    
    \hline
    \end{tabular}
    \end{table*}

\vspace{-1mm}    
\subsection{How does CFIR perform compared to state-of-the-art approaches? (RQ2)}
This section investigates the improved performance facilitated by CFIR under AToMiC base setting by comparing it with OpenCLIP and the D-BEiT-3 model, mainly in text-only fine-tuning settings, because it is a fair comparison to CFIR. We also compare CFIR with whole-model fine-tuned approaches to demonstrate the performance trade-off.

\looseness -1 In the text-only fine-tuning setting, experimental results are illustrated in the bottom block of Table \ref{tab1}. For effectiveness, we observe that CFIR outperforms the previous state-of-the-art models OpenCLIP-Frozen and our proposed D-BEiT-3-Frozen under the AToMiC base across all metrics. Notably, CFIR is more effective with smaller encoder sizes. For instance, CFIR-B outperforms OpenCLIP-B-Frozen, achieving a 0.015 higher score in MRR@10 and an 11.06\% better result in R@1000. Similarly, CFIR-L surpasses OpenCLIP-L-Frozen with a 0.018 increase in MRR@10 and a 5.55\% improvement in R@1000. When compared to D-BEiT-3-B-Frozen, CFIR-B shows a 0.01 improvement in MRR@10 and a significant 7.44\% gain in R@1000. Against D-BEiT-3-L-Frozen, CFIR-L leads in both metrics, showing a 0.016 higher score in MRR@10 and a 4.32\% advance in R@1000. 

In terms of efficiency, compared to prior state-of-the-art MLLM-based methods \cite{cherti2023reproducible,wang2023image}, our approach incurs additional computational time and storage for constructing the entity-based image candidate index and the shared image embedding cache. Specifically, the largest variant of CFIR (CFIR-L) requires an extra 71 GB of storage space and 34 hours of preparation time, which involves building the entity-based image candidate index and the shared image embedding cache. For context, BEiT-3-L requires 2280 hours for a 30-epoch training cycle, the extra 34 hours for CFIR-L's setup is quite minimal which only about 0.14\% of the total time BEiT-3-L needs for training. Furthermore, the integration of an index and cache markedly reduces the training duration for CFIR-L. It only requires 45 hours per epoch, culminating in a total of 930 hours for 30 epochs. Owing to the Decoupling-BEiT-3 architectural efficiency and vector-based distance computation, which involves an entity-based image candidates index for filtering and image embedding cache, CFIR significantly streamlines the retrieval process. In the AToMiC base setting, CFIR-B has managed to reduce the retrieval time significantly, from 363.6 milliseconds to mere 4.2 milliseconds, when compared to D-BEiT-3-B-Frozen. For CFIR-L, the reduction in retrieval time is even more remarkable, the time required for CFIR-L to retrieve images is only about 1.1\% of the time taken by D-BEiT-3-L-Frozen.

When comparing to full-model fine-tuning approaches, our CFIR still outperforms the previous state-of-the-art model OpenCLIP-Full. CFIR-B gains a 6.04\% improvement on Recall@1000 compared to OpenCLIP-B-Full. When compared with the robust whole-model fine-tuned D-BEiT-3-Full, CFIR-B not only slightly surpasses BEiT-3-B-Full in performance but does so with only 56.12\% of its parameters, consequently reducing training time by 68.7\%. In terms of large models, CFIR-L sustains competitive retrieval performance in comparison to D-BEiT-3-L-Full, albeit with a minor decline in Recall@1000, while achieving a 40.7\% reduction in training time due to the substantial reduction in the length of each document summary. When set against full-model fine-tuning methods, CFIR-B exhibits higher gains in Recall@1000 and compared to D-BEiT-3-B-Full, with 0.07\% increase in Recall@1000 in the AToMiC base setting. Similarly, the Recall@1000 performance gap between CFIR-L and D-BEiT-3-L-Full narrows to 1.71\%. This result underscores CFIR's better scalability as the candidate set size increases.

To summarize, under AToMiC base setting, CFIR outperforms state-of-the-art models OpenCLIP and D-BEiT-3 in both effectiveness and efficiency in text-only fine-tuning settings. While it incurs modest additional costs, the efficiency gains in training and retrieval time are significant. Even when compared to whole-model fine-tuning approaches, CFIR maintains a competitive performance while requiring fewer parameters and significantly less training and retrieval time, thereby proving its viability for large-scale applications. This result answers our proposed research question 2 that how does CFIR perform compared to state-of-the-art approaches.
\begin{table*} 
    \centering
    \vspace{-1mm}
    \caption{Ablation studies on AToMic base setting. T-t means training time (Hour/epoch) and R-t means retrieval time (millisecond/query).}
    \vspace{-1mm}
    \label{tab3}
    \small
    \fontsize{9}{12.5}\selectfont
    \renewcommand\tabcolsep{5.0pt}
        \begin{tabular}{c|ccccc|cccc}
        \hline
    
        \hline 
        Method & Cache & Index & Entity & Summary  & Doc & MRR@10   & R@1000 & T-t & R-t \\
        \hline
        \multirow{5}{*}{CFIP-L} & - & - & -     & -   & \checkmark   & 0.065	 & 51.36 & 53    &  425.1  \\  
         
         & \checkmark & \checkmark & \checkmark & -    & -   &  0.006     	     & 13.15    & 0     & 0 \\ 
         & - & - &  -     & \checkmark   & -     & 0.069		& 53.91 & 45     &  425.1  \\
         & \checkmark & \checkmark & \checkmark & -    & \checkmark   &  0.075     	     & 54.61    & 53   &   13.0 \\%
        & \checkmark & \checkmark & \checkmark & \checkmark   & -    &  0.081     	  & 55.68   & 45     & 4.7 \\
        
        \hline
    
        \hline
        \end{tabular} 
    \end{table*}
\subsection{CFIR scalability in AToMiC Large Setting (RQ3)}

In addition to exploring CFIR's performance in the AToMiC base setting, this section extends the analysis to the more demanding AToMiC large Setting that has three times more images and long-text compare to AToMiC base setting. Here, we also investigates CFIR's enhanced capabilities by comparing it with the D-BEiT-3 model in two scenarios: text-only fine-tuning and whole-model fine-tuning. Combining with our result in AToMiC base setting, we provide a comprehensive comparison between our proposed CFIR framework and previous state-of-the-art methods, and we aim to show CFIR's adaptability and robust performance across diverse fine-tuning scenarios.

The experimental results in the text-only fine-tuning setting, as detailed in the bottom block of Table \ref{tab2}, reveal that our proposed CFIR surpasses D-BEiT-3-Frozen in the AToMiC large setting across all evaluated metrics. In particular, CFIR-B shows a notable increase of 0.006 and 1.91\% in MRR@10 and R@1000, respectively, when compared to D-BEiT-3-B-Frozen. Similarly, CFIR-L demonstrates a significant lead with improvements of 0.004 and 3.15\% in MRR@10 and R@1000, respectively, over D-BEiT-3-L-Frozen. Furthermore, this level of efficiency is consistently observed in the more demanding AToMiC large setting, paralleling results from the AToMiC base setting. In terms of retrieval time, CFIR-B manages to cut down the retrieval time by 985 milliseconds per query compared to D-BEiT-3-B-Frozen.  These reductions underscore the effectiveness of CFIR in optimizing the retrieval latency. When set against full-model fine-tuning methods, CFIR-B exhibits higher gains in Recall@1000 and MRR@10 compared to D-BEiT-3-B-Full, with a 0.96\%. Similarly, the Recall@1000 performance gap between CFIR-L and D-BEiT-3-L-Full narrows to 0.86\%. This result underscores CFIR's better scalability as the candidate set size increases.

In conclusion, within the challenging AToMiC large setting, CFIR demonstrates significantly improvement performance over D-BEiT-3 in terms efficiency in both text-only and whole-model fine-tuning scenarios. These findings underscore CFIR's robust capabilities in enhancing retrieval processes, scaling effective and maintain its power in larger datasets.

\section{Analysis}
This section comprehensively evaluates our CFIR framework through three focused analyses. We dissect CFIR's main components in an ablation study, explore Top-K effects, and offer a qualitative analysis of the model's practical utility.

\begin{figure*}
\includegraphics[width=1\textwidth]{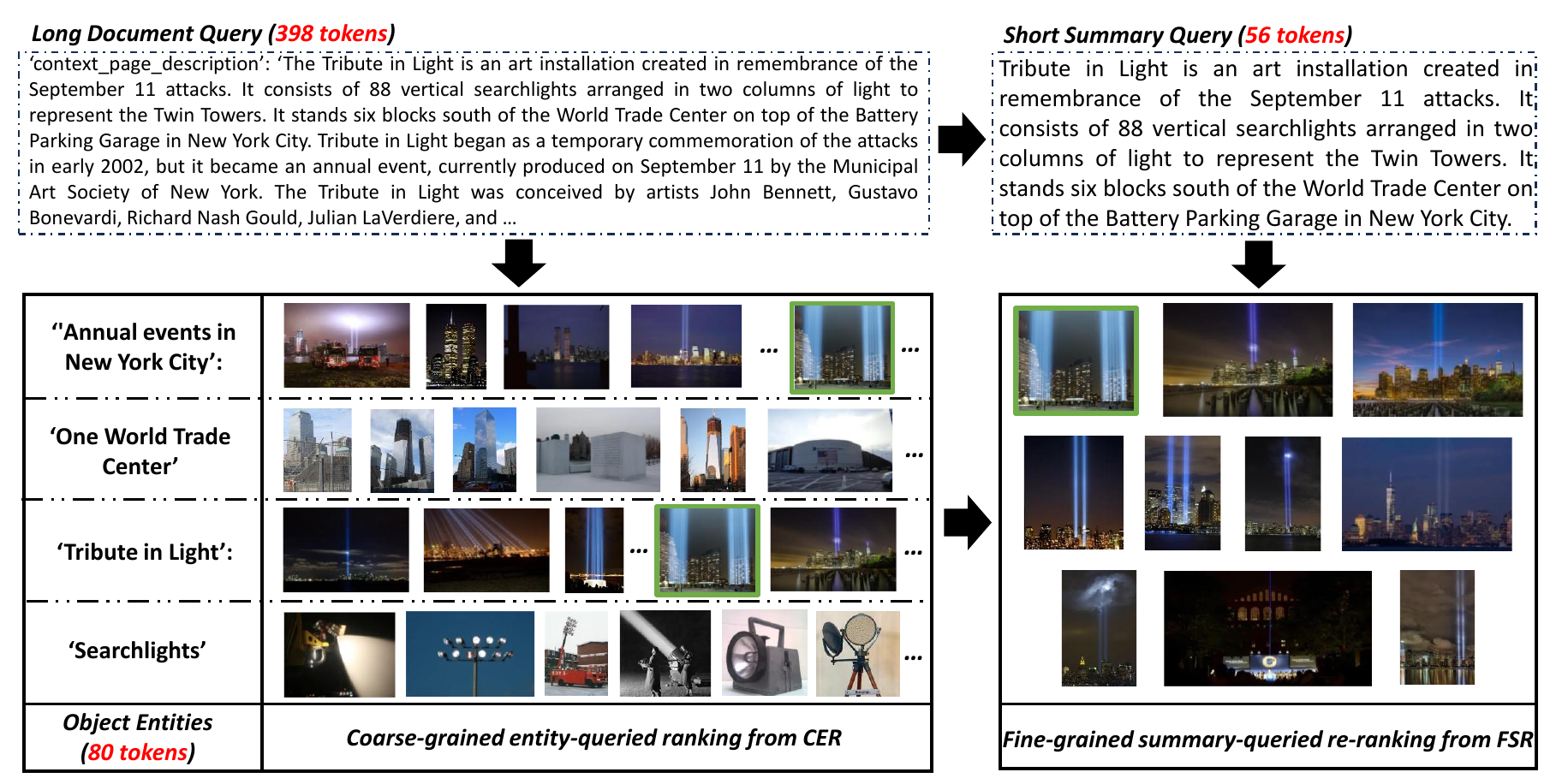}
\caption{An example of our CFIR for long-text image retrieval.} \label{Qualitative}
\end{figure*}

\begin{table} 
    \centering
    \caption{The performance impact of varying Top-K in Entity-based Ranking.}
    \label{tab4}
    \fontsize{9}{13}\selectfont
    \renewcommand\tabcolsep{5.0pt}
        \begin{tabular}{c|c|c|c|c}
        \hline
    
        \hline
        Method & Top-K & Retrieval-time  & MRR@10   & R@1000 \\
        \hline
        \multirow{4}{*}{CFIP-L} & Top-1000  &   1.5 ms/query    & 0.033     & 38.32   \\ 
         & Top-5000     & 2.9 ms/query     &  0.051      & 50.18        \\ 
        \cline{2-5}
        & \textbf{Top-10000} & 4.7 ms/query   &  \textbf{0.081}        & \textbf{55.68}        \\ \cline{2-5} 
        & Top-15000 & 28.8 ms/query   &  0.079       & 55.35        \\ 
        
        \hline
    
        \hline
        \end{tabular}
        \vspace{-3mm}
    \end{table}

\subsection{Ablation Study of CFIR}
We examine the contributions of the Entity-based Ranking (ER) and Summary-based Re-ranking (SR) stages in CFIR, alongside the CFIR-L variant's use of a pre-computed shared index and image embedding cache. We benchmark against state-of-the-art MLLM-based approaches (Row 1, Table \ref{tab3}). Using only the SR stage (Row 3, Table \ref{tab3}) leads to a significant improvement of 2.55\% on R@1000 and reduces training time from 53 to 45 hours per epoch. This indicates that concise summaries enhance both retrieval and training efficiency by reducing document ambiguity.

A comparative analysis (Rows 1 and 4, Table \ref{tab3}) shows that integrating ER, a pre-computed shared index, and image embedding cache results in improved retrieval effectiveness. This integration yields an improvement of 0.01 in MRR@10 and 3.25\% in R@1000, and reduces the retrieval time from 425.1 millisecond to 13 millisecond per query. The completed CFIR-L model (Row 5, Table \ref{tab3}) surpasses all other configurations in both effectiveness and efficiency, requiring significantly less training and retrieval time. Specifically, it cuts down the training time from 53 to 45 hours per epoch and the retrieval time from 425.1 millisecond to 4.7 millisecond per query, while achieving a 0.016 increase in MRR@10 and a 4.32\% increase in R@1000 compared to MLLM-based approaches (Row 1, Table \ref{tab3}).

In summary, the ER and SR stages in CFIR contribute to both efficient image retrieval and improved performance metrics. Incorporating a pre-computed shared index and image embedding cache further reduces retrieval and training time dramatically.
\vspace{-2mm}

\subsection{Performance Impact of Varying Top-K in Entity-based Ranking}
\looseness -1 In this section, we examine the performance implications of varying the value of Top-K in the Entity-based Ranking (ER) Stage, as detailed in Table \ref{tab4}. Our observations indicate that as the value of \(k\) increases, the re-ranking effectiveness of SR initially experiences a significant boost, only to decrease slightly afterward. Specifically, there is an increase of 0.048 on MRR@10 and 17.36\% on R@1000 when Top-K increases from 1000 to 10000. However, this is followed by a slight decrease, from 0.081 to 0.079 in MRR@10 and from 55.68\% to 55.35\% in R@1000. This phenomenon can be attributed to the increased likelihood of identifying images that are relevant to the query document. However, this also results in the inclusion of additional 'interference' candidates—images that are semantically similar but not identical matches, thereby slightly diminishing effectiveness. Concurrently, the retrieval time shows an upward trend as candidate numbers increase.

To strike a balance between retrieval effectiveness and efficiency, we choose the image candidates for each entity at 10,000 for the SR stage. This configuration necessitates a mere 4.7 millisecond per query in the AToMiC base setting; specifically, it yields scores of 0.081 on MRR@10, and 55.68\% on R@1000.

\vspace{-1mm}
\subsection{Qualitative Evaluation of CFIR}

Figure \ref{Qualitative} showcases the practical outcomes of applying our CFIR framework within the AToMiC base setting. The figure presents a comprehensive view of the retrieval process, starting with the original long document, which consists of 398 tokens, and proceeding through the crucial stages of entity extraction and summarization. From the long text, 80 entities are extracted, highlighting the essential elements that facilitate the subsequent retrieval tasks. Furthermore, a brief summary of 56 tokens is generated, shorten the document into a concise query. Additionally, it presents the ranked image results yielded by both the ER and SR stages of our framework.

Due to spatial constraints, we limit our visualization to four randomly selected entities and their associated top-ranked images, as generated by the first ER stage. This choice in visualization serves to underscore ER's efficacy in not only retrieving a set of relevant images based on individual entities but also in effectively filtering out irrelevant images from a large-scale collection. For instance, the entities "Tribute in Light" and "Annual events in New York City" yield numerous relevant image candidates, including the target image. This approach thereby paves the way for the subsequent SR stage to focus on finding the exact matching image of key document information. As a result, CFIR not only enhances the retrieval accuracy but also speeds up the searching process, ensuring that the most contextually relevant images are brought to the forefront for final selection.

\vspace{-1mm}
\section{Conclusion}
In this paper, we address the challenges associated with Large-Scale Long-Text to Image Retrieval (LLIR). To this end, we introduce a novel, two-stage Coarse-to-Fine Index-Shared Retrieval (CFIR) framework designed to mitigate the ambiguity in long documents while optimizing retrieval effectiveness and efficiency. The CFIR framework is modular, comprising two principal stages: Entity-based Ranking (ER) and Summary-based Re-ranking (SR). These stages employ our proposed encoding model, decoupling-BEiT-3, which enables vector-based distance similarity inference and the use of a pre-computed, shared entity-based candidates index and image embedding cache. This significantly improves training and retrieval efficiency. Our experimental results demonstrate that CFIR outperforms existing state-of-the-art methods in the AToMiC LLIR task, as confirmed by both quantitative and qualitative evaluations. These findings have practical implications in various applications that necessitate efficient and effective large-scale image retrieval from long documents.

\clearpage
\bibliographystyle{ACM-Reference-Format}
\bibliography{base}


\begin{thebibliography}{47}


\ifx \showCODEN    \undefined \def \showCODEN     #1{\unskip}     \fi
\ifx \showDOI      \undefined \def \showDOI       #1{#1}\fi
\ifx \showISBNx    \undefined \def \showISBNx     #1{\unskip}     \fi
\ifx \showISBNxiii \undefined \def \showISBNxiii  #1{\unskip}     \fi
\ifx \showISSN     \undefined \def \showISSN      #1{\unskip}     \fi
\ifx \showLCCN     \undefined \def \showLCCN      #1{\unskip}     \fi
\ifx \shownote     \undefined \def \shownote      #1{#1}          \fi
\ifx \showarticletitle \undefined \def \showarticletitle #1{#1}   \fi
\ifx \showURL      \undefined \def \showURL       {\relax}        \fi
\providecommand\bibfield[2]{#2}
\providecommand\bibinfo[2]{#2}
\providecommand\natexlab[1]{#1}
\providecommand\showeprint[2][]{arXiv:#2}

\bibitem[Chen et~al\mbox{.}(2020)]%
        {IMRAM}
\bibfield{author}{\bibinfo{person}{Hui Chen}, \bibinfo{person}{Guiguang Ding}, \bibinfo{person}{Xudong Liu}, \bibinfo{person}{Zijia Lin}, \bibinfo{person}{Ji Liu}, {and} \bibinfo{person}{Jungong Han}.} \bibinfo{year}{2020}\natexlab{}.
\newblock \showarticletitle{Imram: Iterative matching with recurrent attention memory for cross-modal image-text retrieval}. In \bibinfo{booktitle}{\emph{CVPR}}. \bibinfo{pages}{12655--12663}.
\newblock


\bibitem[Cherti et~al\mbox{.}(2023)]%
        {cherti2023reproducible}
\bibfield{author}{\bibinfo{person}{Mehdi Cherti}, \bibinfo{person}{Romain Beaumont}, \bibinfo{person}{Ross Wightman}, \bibinfo{person}{Mitchell Wortsman}, \bibinfo{person}{Gabriel Ilharco}, \bibinfo{person}{Cade Gordon}, \bibinfo{person}{Christoph Schuhmann}, \bibinfo{person}{Ludwig Schmidt}, {and} \bibinfo{person}{Jenia Jitsev}.} \bibinfo{year}{2023}\natexlab{}.
\newblock \showarticletitle{Reproducible scaling laws for contrastive language-image learning}. In \bibinfo{booktitle}{\emph{CVPR}}. \bibinfo{pages}{2818--2829}.
\newblock


\bibitem[Cubuk et~al\mbox{.}(2019)]%
        {DBLP:conf/cvpr/CubukZMVL19}
\bibfield{author}{\bibinfo{person}{Ekin~D Cubuk}, \bibinfo{person}{Barret Zoph}, \bibinfo{person}{Dandelion Mane}, \bibinfo{person}{Vijay Vasudevan}, {and} \bibinfo{person}{Quoc~V Le}.} \bibinfo{year}{2019}\natexlab{}.
\newblock \showarticletitle{Autoaugment: Learning augmentation strategies from data}. In \bibinfo{booktitle}{\emph{CVPR}}. \bibinfo{pages}{113--123}.
\newblock


\bibitem[Deldjoo et~al\mbox{.}(2020)]%
        {deldjoo2020recommender}
\bibfield{author}{\bibinfo{person}{Yashar Deldjoo}, \bibinfo{person}{Markus Schedl}, \bibinfo{person}{Paolo Cremonesi}, {and} \bibinfo{person}{Gabriella Pasi}.} \bibinfo{year}{2020}\natexlab{}.
\newblock \showarticletitle{Recommender systems leveraging multimedia content}.
\newblock \bibinfo{journal}{\emph{ACM CSUR}} \bibinfo{volume}{53}, \bibinfo{number}{5} (\bibinfo{year}{2020}), \bibinfo{pages}{1--38}.
\newblock


\bibitem[Devlin et~al\mbox{.}(2019)]%
        {devlin2019bert}
\bibfield{author}{\bibinfo{person}{Jacob Devlin}, \bibinfo{person}{Ming-Wei Chang}, \bibinfo{person}{Kenton Lee}, {and} \bibinfo{person}{Kristina Toutanova}.} \bibinfo{year}{2019}\natexlab{}.
\newblock \showarticletitle{Bert: Pre-training of deep bidirectional transformers for language understanding}.
\newblock \bibinfo{journal}{\emph{NAACL}} (\bibinfo{year}{2019}).
\newblock


\bibitem[Feng et~al\mbox{.}(2014)]%
        {feng2014cross}
\bibfield{author}{\bibinfo{person}{Fangxiang Feng}, \bibinfo{person}{Xiaojie Wang}, {and} \bibinfo{person}{Ruifan Li}.} \bibinfo{year}{2014}\natexlab{}.
\newblock \showarticletitle{Cross-modal retrieval with correspondence autoencoder}. In \bibinfo{booktitle}{\emph{ACM MM}}. \bibinfo{pages}{7--16}.
\newblock


\bibitem[Ge et~al\mbox{.}(2021)]%
        {ge2021structured}
\bibfield{author}{\bibinfo{person}{Xuri Ge}, \bibinfo{person}{Fuhai Chen}, \bibinfo{person}{Joemon~M Jose}, \bibinfo{person}{Zhilong Ji}, \bibinfo{person}{Zhongqin Wu}, {and} \bibinfo{person}{Xiao Liu}.} \bibinfo{year}{2021}\natexlab{}.
\newblock \showarticletitle{Structured multi-modal feature embedding and alignment for image-sentence retrieval}. In \bibinfo{booktitle}{\emph{ACM MM}}. \bibinfo{pages}{5185--5193}.
\newblock


\bibitem[Ge et~al\mbox{.}(2023)]%
        {ge2023cross}
\bibfield{author}{\bibinfo{person}{Xuri Ge}, \bibinfo{person}{Fuhai Chen}, \bibinfo{person}{Songpei Xu}, \bibinfo{person}{Fuxiang Tao}, {and} \bibinfo{person}{Joemon~M Jose}.} \bibinfo{year}{2023}\natexlab{}.
\newblock \showarticletitle{Cross-modal Semantic Enhanced Interaction for Image-Sentence Retrieval}. In \bibinfo{booktitle}{\emph{Proceedings of the IEEE/CVF Winter Conference on Applications of Computer Vision}}. \bibinfo{pages}{1022--1031}.
\newblock


\bibitem[Ge et~al\mbox{.}(2024)]%
        {ge20243shnet}
\bibfield{author}{\bibinfo{person}{Xuri Ge}, \bibinfo{person}{Songpei Xu}, \bibinfo{person}{Fuhai Chen}, \bibinfo{person}{Jie Wang}, \bibinfo{person}{Guoxin Wang}, \bibinfo{person}{Shan An}, {and} \bibinfo{person}{Joemon~M Jose}.} \bibinfo{year}{2024}\natexlab{}.
\newblock \showarticletitle{3SHNet: Boosting image--sentence retrieval via visual semantic--spatial self-highlighting}.
\newblock \bibinfo{journal}{\emph{Information Processing \& Management}} \bibinfo{volume}{61}, \bibinfo{number}{4} (\bibinfo{year}{2024}), \bibinfo{pages}{103716}.
\newblock


\bibitem[Hong et~al\mbox{.}(2021)]%
        {DBLP:conf/sigir/HongJLWCC21}
\bibfield{author}{\bibinfo{person}{Weixiang Hong}, \bibinfo{person}{Kaixiang Ji}, \bibinfo{person}{Jiajia Liu}, \bibinfo{person}{Jian Wang}, \bibinfo{person}{Jingdong Chen}, {and} \bibinfo{person}{Wei Chu}.} \bibinfo{year}{2021}\natexlab{}.
\newblock \showarticletitle{Gilbert: Generative vision-language pre-training for image-text retrieval}. In \bibinfo{booktitle}{\emph{ACM SIGIR}}. \bibinfo{pages}{1379--1388}.
\newblock


\bibitem[Hu et~al\mbox{.}(2023)]%
        {DBLP:conf/sigir/HuGCWY23}
\bibfield{author}{\bibinfo{person}{Xuming Hu}, \bibinfo{person}{Zhijiang Guo}, \bibinfo{person}{Junzhe Chen}, \bibinfo{person}{Lijie Wen}, {and} \bibinfo{person}{Philip~S. Yu}.} \bibinfo{year}{2023}\natexlab{}.
\newblock \showarticletitle{{MR2:} {A} Benchmark for Multimodal Retrieval-Augmented Rumor Detection in Social Media}. In \bibinfo{booktitle}{\emph{Proceedings of the 46th International {ACM} {SIGIR} Conference on Research and Development in Information Retrieval, {SIGIR} 2023, Taipei, Taiwan, July 23-27, 2023}}, \bibfield{editor}{\bibinfo{person}{Hsin{-}Hsi Chen}, \bibinfo{person}{Wei{-}Jou~(Edward) Duh}, \bibinfo{person}{Hen{-}Hsen Huang}, \bibinfo{person}{Makoto~P. Kato}, \bibinfo{person}{Josiane Mothe}, {and} \bibinfo{person}{Barbara Poblete}} (Eds.). \bibinfo{publisher}{{ACM}}, \bibinfo{pages}{2901--2912}.
\newblock
\urldef\tempurl%
\url{https://doi.org/10.1145/3539618.3591896}
\showDOI{\tempurl}


\bibitem[Huang et~al\mbox{.}(2018)]%
        {huang2018learning}
\bibfield{author}{\bibinfo{person}{Yan Huang}, \bibinfo{person}{Qi Wu}, \bibinfo{person}{Chunfeng Song}, {and} \bibinfo{person}{Liang Wang}.} \bibinfo{year}{2018}\natexlab{}.
\newblock \showarticletitle{Learning semantic concepts and order for image and sentence matching}. In \bibinfo{booktitle}{\emph{CVPR}}. \bibinfo{pages}{6163--6171}.
\newblock


\bibitem[Kennedy et~al\mbox{.}(2005)]%
        {kennedy2005automatic}
\bibfield{author}{\bibinfo{person}{Lyndon~S Kennedy}, \bibinfo{person}{Apostol Natsev}, {and} \bibinfo{person}{Shih-Fu Chang}.} \bibinfo{year}{2005}\natexlab{}.
\newblock \showarticletitle{Automatic discovery of query-class-dependent models for multimodal search}. In \bibinfo{booktitle}{\emph{ACM MM}}. \bibinfo{pages}{882--891}.
\newblock


\bibitem[Lewis et~al\mbox{.}(2019)]%
        {lewis2020bart}
\bibfield{author}{\bibinfo{person}{Mike Lewis}, \bibinfo{person}{Yinhan Liu}, \bibinfo{person}{Naman Goyal}, \bibinfo{person}{Marjan Ghazvininejad}, \bibinfo{person}{Abdelrahman Mohamed}, \bibinfo{person}{Omer Levy}, \bibinfo{person}{Ves Stoyanov}, {and} \bibinfo{person}{Luke Zettlemoyer}.} \bibinfo{year}{2019}\natexlab{}.
\newblock \showarticletitle{Bart: Denoising sequence-to-sequence pre-training for natural language generation, translation, and comprehension}.
\newblock \bibinfo{journal}{\emph{ACL}} (\bibinfo{year}{2019}).
\newblock


\bibitem[Li et~al\mbox{.}(2020a)]%
        {DBLP:journals/corr/abs-1908-06066}
\bibfield{author}{\bibinfo{person}{Gen Li}, \bibinfo{person}{Nan Duan}, \bibinfo{person}{Yuejian Fang}, \bibinfo{person}{Ming Gong}, {and} \bibinfo{person}{Daxin Jiang}.} \bibinfo{year}{2020}\natexlab{a}.
\newblock \showarticletitle{Unicoder-vl: A universal encoder for vision and language by cross-modal pre-training}. In \bibinfo{booktitle}{\emph{AAAI}}, Vol.~\bibinfo{volume}{34}. \bibinfo{pages}{11336--11344}.
\newblock


\bibitem[Li et~al\mbox{.}(2022)]%
        {li2022blip}
\bibfield{author}{\bibinfo{person}{Junnan Li}, \bibinfo{person}{Dongxu Li}, \bibinfo{person}{Caiming Xiong}, {and} \bibinfo{person}{Steven Hoi}.} \bibinfo{year}{2022}\natexlab{}.
\newblock \showarticletitle{Blip: Bootstrapping language-image pre-training for unified vision-language understanding and generation}. In \bibinfo{booktitle}{\emph{ICML}}. PMLR, \bibinfo{pages}{12888--12900}.
\newblock


\bibitem[Li et~al\mbox{.}(2019b)]%
        {li2019visual}
\bibfield{author}{\bibinfo{person}{Kunpeng Li}, \bibinfo{person}{Yulun Zhang}, \bibinfo{person}{Kai Li}, \bibinfo{person}{Yuanyuan Li}, {and} \bibinfo{person}{Yun Fu}.} \bibinfo{year}{2019}\natexlab{b}.
\newblock \showarticletitle{Visual semantic reasoning for image-text matching}. In \bibinfo{booktitle}{\emph{ICCV}}. \bibinfo{pages}{4654--4662}.
\newblock


\bibitem[Li et~al\mbox{.}(2019a)]%
        {DBLP:journals/corr/abs-1908-03557}
\bibfield{author}{\bibinfo{person}{Liunian~Harold Li}, \bibinfo{person}{Mark Yatskar}, \bibinfo{person}{Da Yin}, \bibinfo{person}{Cho{-}Jui Hsieh}, {and} \bibinfo{person}{Kai{-}Wei Chang}.} \bibinfo{year}{2019}\natexlab{a}.
\newblock \showarticletitle{VisualBERT: {A} Simple and Performant Baseline for Vision and Language}.
\newblock \bibinfo{journal}{\emph{CoRR}}  \bibinfo{volume}{abs/1908.03557} (\bibinfo{year}{2019}).
\newblock


\bibitem[Li et~al\mbox{.}(2020b)]%
        {DBLP:conf/eccv/Li0LZHZWH0WCG20}
\bibfield{author}{\bibinfo{person}{Xiujun Li}, \bibinfo{person}{Xi Yin}, \bibinfo{person}{Chunyuan Li}, \bibinfo{person}{Pengchuan Zhang}, \bibinfo{person}{Xiaowei Hu}, \bibinfo{person}{Lei Zhang}, \bibinfo{person}{Lijuan Wang}, \bibinfo{person}{Houdong Hu}, \bibinfo{person}{Li Dong}, \bibinfo{person}{Furu Wei}, {et~al\mbox{.}}} \bibinfo{year}{2020}\natexlab{b}.
\newblock \showarticletitle{Oscar: Object-semantics aligned pre-training for vision-language tasks}. In \bibinfo{booktitle}{\emph{ECCV}}. Springer, \bibinfo{pages}{121--137}.
\newblock


\bibitem[Lin et~al\mbox{.}(2023)]%
        {DBLP:conf/sigir/LinJSLSN23}
\bibfield{author}{\bibinfo{person}{Dengtian Lin}, \bibinfo{person}{Liqiang Jing}, \bibinfo{person}{Xuemeng Song}, \bibinfo{person}{Meng Liu}, \bibinfo{person}{Teng Sun}, {and} \bibinfo{person}{Liqiang Nie}.} \bibinfo{year}{2023}\natexlab{}.
\newblock \showarticletitle{Adapting Generative Pretrained Language Model for Open-domain Multimodal Sentence Summarization}. In \bibinfo{booktitle}{\emph{Proceedings of the 46th International {ACM} {SIGIR} Conference on Research and Development in Information Retrieval, {SIGIR} 2023, Taipei, Taiwan, July 23-27, 2023}}, \bibfield{editor}{\bibinfo{person}{Hsin{-}Hsi Chen}, \bibinfo{person}{Wei{-}Jou~(Edward) Duh}, \bibinfo{person}{Hen{-}Hsen Huang}, \bibinfo{person}{Makoto~P. Kato}, \bibinfo{person}{Josiane Mothe}, {and} \bibinfo{person}{Barbara Poblete}} (Eds.). \bibinfo{publisher}{{ACM}}, \bibinfo{pages}{195--204}.
\newblock
\urldef\tempurl%
\url{https://doi.org/10.1145/3539618.3591633}
\showDOI{\tempurl}


\bibitem[Lin et~al\mbox{.}(2014)]%
        {lin2014microsoft}
\bibfield{author}{\bibinfo{person}{Tsung-Yi Lin}, \bibinfo{person}{Michael Maire}, \bibinfo{person}{Serge Belongie}, \bibinfo{person}{James Hays}, \bibinfo{person}{Pietro Perona}, \bibinfo{person}{Deva Ramanan}, \bibinfo{person}{Piotr Doll{\'a}r}, {and} \bibinfo{person}{C~Lawrence Zitnick}.} \bibinfo{year}{2014}\natexlab{}.
\newblock \showarticletitle{Microsoft coco: Common objects in context}. In \bibinfo{booktitle}{\emph{ECCV}}. Springer, \bibinfo{pages}{740--755}.
\newblock


\bibitem[Long et~al\mbox{.}(2022)]%
        {Gradual}
\bibfield{author}{\bibinfo{person}{Siqu Long}, \bibinfo{person}{Soyeon~Caren Han}, \bibinfo{person}{Xiaojun Wan}, {and} \bibinfo{person}{Josiah Poon}.} \bibinfo{year}{2022}\natexlab{}.
\newblock \showarticletitle{Gradual: Graph-based dual-modal representation for image-text matching}. In \bibinfo{booktitle}{\emph{WACV}}. \bibinfo{pages}{3459--3468}.
\newblock


\bibitem[Long et~al\mbox{.}(2023a)]%
        {long2023robollm}
\bibfield{author}{\bibinfo{person}{Zijun Long}, \bibinfo{person}{George Killick}, \bibinfo{person}{Richard McCreadie}, {and} \bibinfo{person}{Gerardo~Aragon Camarasa}.} \bibinfo{year}{2023}\natexlab{a}.
\newblock \showarticletitle{Robollm: Robotic vision tasks grounded on multimodal large language models}.
\newblock \bibinfo{journal}{\emph{arXiv preprint arXiv:2310.10221}} (\bibinfo{year}{2023}).
\newblock


\bibitem[Long et~al\mbox{.}(2024)]%
        {long2024multiway}
\bibfield{author}{\bibinfo{person}{Zijun Long}, \bibinfo{person}{George Killick}, \bibinfo{person}{Richard McCreadie}, {and} \bibinfo{person}{Gerardo~Aragon Camarasa}.} \bibinfo{year}{2024}\natexlab{}.
\newblock \showarticletitle{Multiway-Adapter: Adapting Multimodal Large Language Models for Scalable Image-Text Retrieval}. In \bibinfo{booktitle}{\emph{ICASSP 2024-2024 IEEE International Conference on Acoustics, Speech and Signal Processing (ICASSP)}}. IEEE, \bibinfo{pages}{6580--6584}.
\newblock


\bibitem[Long et~al\mbox{.}(2023b)]%
        {RN152}
\bibfield{author}{\bibinfo{person}{Zijun Long}, \bibinfo{person}{George Killick}, \bibinfo{person}{Richard McCreadie}, \bibinfo{person}{Gerardo~Aragon Camarasa}, {and} \bibinfo{person}{Zaiqiao Meng}.} \bibinfo{year}{2023}\natexlab{b}.
\newblock \showarticletitle{When hard negative sampling meets supervised contrastive learning}.
\newblock \bibinfo{journal}{\emph{arXiv preprint arXiv:2308.14893}}.
\newblock


\bibitem[Long et~al\mbox{.}(2023c)]%
        {RN156}
\bibfield{author}{\bibinfo{person}{Zijun Long}, \bibinfo{person}{George Killick}, \bibinfo{person}{Lipeng Zhuang}, \bibinfo{person}{Richard McCreadie}, \bibinfo{person}{Gerardo~Aragon Camarasa}, {and} \bibinfo{person}{Paul Henderson}.} \bibinfo{year}{2023}\natexlab{c}.
\newblock \showarticletitle{Elucidating and overcoming the challenges of label noise in supervised contrastive learning}.
\newblock \bibinfo{journal}{\emph{arXiv preprint arXiv:2311.16481}}.
\newblock


\bibitem[Long and McCreadie({[n.\,d.]})]%
        {RN150}
\bibfield{author}{\bibinfo{person}{Zijun Long} {and} \bibinfo{person}{Richard McCreadie}.} \bibinfo{year}{[n.\,d.]}\natexlab{}.
\newblock \showarticletitle{Is Multi-Modal Data Key for Crisis Content Categorization on Social Media?}. In \bibinfo{booktitle}{\emph{19th International Conference on Information Systems for Crisis Response and Management (ISCRAM 2022)}}.
\newblock


\bibitem[Long et~al\mbox{.}({[n.\,d.]})]%
        {long157}
\bibfield{author}{\bibinfo{person}{Zijun Long}, \bibinfo{person}{Richard McCreadie}, \bibinfo{person}{Gerardo Aragon~Camarasa}, {and} \bibinfo{person}{Zaiqiao Meng}.} \bibinfo{year}{[n.\,d.]}\natexlab{}.
\newblock \showarticletitle{LACVIT: A Label-aware Contrastive Fine-tuning Framework for Vision Transformers}. In \bibinfo{booktitle}{\emph{IEEE International Conference on Acoustics, Speech, and Signal Processing (ICASSP 2024)}}.
\newblock


\bibitem[Long et~al\mbox{.}(2023d)]%
        {long2024crisisvit}
\bibfield{author}{\bibinfo{person}{Zijun Long}, \bibinfo{person}{Richard McCreadie}, {and} \bibinfo{person}{Muhammad Imran}.} \bibinfo{year}{2023}\natexlab{d}.
\newblock \showarticletitle{Crisisvit: A robust vision transformer for crisis image classification}.
\newblock \bibinfo{journal}{\emph{arXiv preprint arXiv:2401.02838}} (\bibinfo{year}{2023}).
\newblock


\bibitem[Lu et~al\mbox{.}(2019)]%
        {DBLP:conf/nips/LuBPL19}
\bibfield{author}{\bibinfo{person}{Jiasen Lu}, \bibinfo{person}{Dhruv Batra}, \bibinfo{person}{Devi Parikh}, {and} \bibinfo{person}{Stefan Lee}.} \bibinfo{year}{2019}\natexlab{}.
\newblock \showarticletitle{Vilbert: Pretraining task-agnostic visiolinguistic representations for vision-and-language tasks}.
\newblock \bibinfo{journal}{\emph{NeurIPS}}  \bibinfo{volume}{32} (\bibinfo{year}{2019}).
\newblock


\bibitem[Qu et~al\mbox{.}(2023)]%
        {qu2023learnable}
\bibfield{author}{\bibinfo{person}{Leigang Qu}, \bibinfo{person}{Meng Liu}, \bibinfo{person}{Wenjie Wang}, \bibinfo{person}{Zhedong Zheng}, \bibinfo{person}{Liqiang Nie}, {and} \bibinfo{person}{Tat-Seng Chua}.} \bibinfo{year}{2023}\natexlab{}.
\newblock \showarticletitle{Learnable Pillar-based Re-ranking for Image-Text Retrieval}.
\newblock \bibinfo{journal}{\emph{ACM SIGIR}} (\bibinfo{year}{2023}).
\newblock


\bibitem[Radford et~al\mbox{.}(2021)]%
        {radford2021learning}
\bibfield{author}{\bibinfo{person}{Alec Radford}, \bibinfo{person}{Jong~Wook Kim}, \bibinfo{person}{Chris Hallacy}, \bibinfo{person}{Aditya Ramesh}, \bibinfo{person}{Gabriel Goh}, \bibinfo{person}{Sandhini Agarwal}, \bibinfo{person}{Girish Sastry}, \bibinfo{person}{Amanda Askell}, \bibinfo{person}{Pamela Mishkin}, \bibinfo{person}{Jack Clark}, {et~al\mbox{.}}} \bibinfo{year}{2021}\natexlab{}.
\newblock \showarticletitle{Learning transferable visual models from natural language supervision}. In \bibinfo{booktitle}{\emph{ICML}}. PMLR, \bibinfo{pages}{8748--8763}.
\newblock


\bibitem[Shao et~al\mbox{.}(2019)]%
        {shao2019two}
\bibfield{author}{\bibinfo{person}{Jie Shao}, \bibinfo{person}{Zhicheng Zhao}, {and} \bibinfo{person}{Fei Su}.} \bibinfo{year}{2019}\natexlab{}.
\newblock \showarticletitle{Two-stage deep learning for supervised cross-modal retrieval}.
\newblock \bibinfo{journal}{\emph{Multimedia Tools and Applications}}  \bibinfo{volume}{78} (\bibinfo{year}{2019}), \bibinfo{pages}{16615--16631}.
\newblock


\bibitem[Singh et~al\mbox{.}(2022)]%
        {singh2022flava}
\bibfield{author}{\bibinfo{person}{Amanpreet Singh}, \bibinfo{person}{Ronghang Hu}, \bibinfo{person}{Vedanuj Goswami}, \bibinfo{person}{Guillaume Couairon}, \bibinfo{person}{Wojciech Galuba}, \bibinfo{person}{Marcus Rohrbach}, {and} \bibinfo{person}{Douwe Kiela}.} \bibinfo{year}{2022}\natexlab{}.
\newblock \showarticletitle{Flava: A foundational language and vision alignment model}. In \bibinfo{booktitle}{\emph{CVPR}}. \bibinfo{pages}{15638--15650}.
\newblock


\bibitem[Song and Soleymani(2019)]%
        {song2019polysemous}
\bibfield{author}{\bibinfo{person}{Yale Song} {and} \bibinfo{person}{Mohammad Soleymani}.} \bibinfo{year}{2019}\natexlab{}.
\newblock \showarticletitle{Polysemous visual-semantic embedding for cross-modal retrieval}. In \bibinfo{booktitle}{\emph{CVPR}}. \bibinfo{pages}{1979--1988}.
\newblock


\bibitem[Srinivasan et~al\mbox{.}(2021)]%
        {srinivasan2021wit}
\bibfield{author}{\bibinfo{person}{Krishna Srinivasan}, \bibinfo{person}{Karthik Raman}, \bibinfo{person}{Jiecao Chen}, \bibinfo{person}{Michael Bendersky}, {and} \bibinfo{person}{Marc Najork}.} \bibinfo{year}{2021}\natexlab{}.
\newblock \showarticletitle{Wit: Wikipedia-based image text dataset for multimodal multilingual machine learning}. In \bibinfo{booktitle}{\emph{ACM SIGIR}}. \bibinfo{pages}{2443--2449}.
\newblock


\bibitem[Tan and Bansal(2019)]%
        {DBLP:conf/emnlp/TanB19}
\bibfield{author}{\bibinfo{person}{Hao Tan} {and} \bibinfo{person}{Mohit Bansal}.} \bibinfo{year}{2019}\natexlab{}.
\newblock \showarticletitle{Lxmert: Learning cross-modality encoder representations from transformers}. In \bibinfo{booktitle}{\emph{EMNLP-IJCNLP}}. \bibinfo{pages}{5099--5110}.
\newblock


\bibitem[Tian et~al\mbox{.}(2022)]%
        {DBLP:conf/sigir/TianWXCSS22}
\bibfield{author}{\bibinfo{person}{Jialin Tian}, \bibinfo{person}{Kai Wang}, \bibinfo{person}{Xing Xu}, \bibinfo{person}{Zuo Cao}, \bibinfo{person}{Fumin Shen}, {and} \bibinfo{person}{Heng~Tao Shen}.} \bibinfo{year}{2022}\natexlab{}.
\newblock \showarticletitle{Multimodal Disentanglement Variational AutoEncoders for Zero-Shot Cross-Modal Retrieval}. In \bibinfo{booktitle}{\emph{{SIGIR} '22: The 45th International {ACM} {SIGIR} Conference on Research and Development in Information Retrieval, Madrid, Spain, July 11 - 15, 2022}}, \bibfield{editor}{\bibinfo{person}{Enrique Amig{\'{o}}}, \bibinfo{person}{Pablo Castells}, \bibinfo{person}{Julio Gonzalo}, \bibinfo{person}{Ben Carterette}, \bibinfo{person}{J.~Shane Culpepper}, {and} \bibinfo{person}{Gabriella Kazai}} (Eds.). \bibinfo{publisher}{{ACM}}, \bibinfo{pages}{960--969}.
\newblock
\urldef\tempurl%
\url{https://doi.org/10.1145/3477495.3532028}
\showDOI{\tempurl}


\bibitem[Wang et~al\mbox{.}(2018)]%
        {wang2018joint}
\bibfield{author}{\bibinfo{person}{Shuhui Wang}, \bibinfo{person}{Yangyu Chen}, \bibinfo{person}{Junbao Zhuo}, \bibinfo{person}{Qingming Huang}, {and} \bibinfo{person}{Qi Tian}.} \bibinfo{year}{2018}\natexlab{}.
\newblock \showarticletitle{Joint global and co-attentive representation learning for image-sentence retrieval}. In \bibinfo{booktitle}{\emph{ACM MM}}. \bibinfo{pages}{1398--1406}.
\newblock


\bibitem[Wang et~al\mbox{.}(2019)]%
        {wang2019matching}
\bibfield{author}{\bibinfo{person}{Tan Wang}, \bibinfo{person}{Xing Xu}, \bibinfo{person}{Yang Yang}, \bibinfo{person}{Alan Hanjalic}, \bibinfo{person}{Heng~Tao Shen}, {and} \bibinfo{person}{Jingkuan Song}.} \bibinfo{year}{2019}\natexlab{}.
\newblock \showarticletitle{Matching images and text with multi-modal tensor fusion and re-ranking}. In \bibinfo{booktitle}{\emph{ACM MM}}. \bibinfo{pages}{12--20}.
\newblock


\bibitem[Wang et~al\mbox{.}(2023)]%
        {wang2023image}
\bibfield{author}{\bibinfo{person}{Wenhui Wang}, \bibinfo{person}{Hangbo Bao}, \bibinfo{person}{Li Dong}, \bibinfo{person}{Johan Bjorck}, \bibinfo{person}{Zhiliang Peng}, \bibinfo{person}{Qiang Liu}, \bibinfo{person}{Kriti Aggarwal}, \bibinfo{person}{Owais~Khan Mohammed}, \bibinfo{person}{Saksham Singhal}, \bibinfo{person}{Subhojit Som}, {et~al\mbox{.}}} \bibinfo{year}{2023}\natexlab{}.
\newblock \showarticletitle{Image as a Foreign Language: BEiT Pretraining for Vision and Vision-Language Tasks}. In \bibinfo{booktitle}{\emph{CVPR}}. \bibinfo{pages}{19175--19186}.
\newblock


\bibitem[Wu et~al\mbox{.}(2021)]%
        {wu2021partially}
\bibfield{author}{\bibinfo{person}{Yaxiong Wu}, \bibinfo{person}{Craig Macdonald}, {and} \bibinfo{person}{Iadh Ounis}.} \bibinfo{year}{2021}\natexlab{}.
\newblock \showarticletitle{Partially observable reinforcement learning for dialog-based interactive recommendation}. In \bibinfo{booktitle}{\emph{ACM RecSys}}. \bibinfo{pages}{241--251}.
\newblock


\bibitem[Yang et~al\mbox{.}(2023)]%
        {yang2023atomic}
\bibfield{author}{\bibinfo{person}{Jheng-Hong Yang}, \bibinfo{person}{Carlos Lassance}, \bibinfo{person}{Rafael Sampaio De~Rezende}, \bibinfo{person}{Krishna Srinivasan}, \bibinfo{person}{Miriam Redi}, \bibinfo{person}{St{\'e}phane Clinchant}, {and} \bibinfo{person}{Jimmy Lin}.} \bibinfo{year}{2023}\natexlab{}.
\newblock \showarticletitle{AToMiC: An Image/Text Retrieval Test Collection to Support Multimedia Content Creation}. In \bibinfo{booktitle}{\emph{ACM SIGIR}}. \bibinfo{pages}{2975--2984}.
\newblock


\bibitem[Yi et~al\mbox{.}(2023)]%
        {RN155}
\bibfield{author}{\bibinfo{person}{Zixuan Yi}, \bibinfo{person}{Zijun Long}, \bibinfo{person}{Iadh Ounis}, \bibinfo{person}{Craig Macdonald}, {and} \bibinfo{person}{Richard Mccreadie}.} \bibinfo{year}{2023}\natexlab{}.
\newblock \showarticletitle{Large multi-modal encoders for recommendation}.
\newblock \bibinfo{journal}{\emph{arXiv preprint arXiv:2310.20343}}.
\newblock


\bibitem[Yoshitaka and Ichikawa(1999)]%
        {yoshitaka1999survey}
\bibfield{author}{\bibinfo{person}{Atsuo Yoshitaka} {and} \bibinfo{person}{Tadao Ichikawa}.} \bibinfo{year}{1999}\natexlab{}.
\newblock \showarticletitle{A survey on content-based retrieval for multimedia databases}.
\newblock \bibinfo{journal}{\emph{IEEE TKDE}} \bibinfo{volume}{11}, \bibinfo{number}{1} (\bibinfo{year}{1999}), \bibinfo{pages}{81--93}.
\newblock


\bibitem[Young et~al\mbox{.}(2014)]%
        {young2014f30k}
\bibfield{author}{\bibinfo{person}{Peter Young}, \bibinfo{person}{Alice Lai}, \bibinfo{person}{Micah Hodosh}, {and} \bibinfo{person}{Julia Hockenmaier}.} \bibinfo{year}{2014}\natexlab{}.
\newblock \showarticletitle{From image descriptions to visual denotations: New similarity metrics for semantic inference over event descriptions}.
\newblock \bibinfo{journal}{\emph{Trans. Assoc. Comput. Linguist.}}  \bibinfo{volume}{2} (\bibinfo{year}{2014}), \bibinfo{pages}{67--78}.
\newblock


\bibitem[Zhao et~al\mbox{.}(2023)]%
        {DBLP:conf/sigir/ZhaoGH0YL23}
\bibfield{author}{\bibinfo{person}{Zijia Zhao}, \bibinfo{person}{Longteng Guo}, \bibinfo{person}{Xingjian He}, \bibinfo{person}{Shuai Shao}, \bibinfo{person}{Zehuan Yuan}, {and} \bibinfo{person}{Jing Liu}.} \bibinfo{year}{2023}\natexlab{}.
\newblock \showarticletitle{{MAMO:} Fine-Grained Vision-Language Representations Learning with Masked Multimodal Modeling}. In \bibinfo{booktitle}{\emph{Proceedings of the 46th International {ACM} {SIGIR} Conference on Research and Development in Information Retrieval, {SIGIR} 2023, Taipei, Taiwan, July 23-27, 2023}}, \bibfield{editor}{\bibinfo{person}{Hsin{-}Hsi Chen}, \bibinfo{person}{Wei{-}Jou~(Edward) Duh}, \bibinfo{person}{Hen{-}Hsen Huang}, \bibinfo{person}{Makoto~P. Kato}, \bibinfo{person}{Josiane Mothe}, {and} \bibinfo{person}{Barbara Poblete}} (Eds.). \bibinfo{publisher}{{ACM}}, \bibinfo{pages}{1528--1538}.
\newblock
\urldef\tempurl%
\url{https://doi.org/10.1145/3539618.3591721}
\showDOI{\tempurl}


\end{thebibliography}

\end{document}